\documentclass[twocolappendix,twocolumn]{aastex631}

\shorttitle{Warm Molecular Gas in NGC~4418 Traced with CO Rovibrational Absorptions}

\shortauthors{Ohyama et al.}

\begin{document}

\title{Warm Molecular Gas in the Central Parsecs of the Buried Nucleus of NGC~4418 Traced with the Fundamental CO Rovibrational Absorptions
\footnote{This research is based on data collected at the Subaru Telescope, which is operated by the National Astronomical Observatory of Japan. We are honored and grateful for the opportunity of observing the Universe from Maunakea, which has the cultural, historical, and natural significance in Hawaii.}}
\author[0000-0001-9490-3582]{Youichi Ohyama}
\affiliation{Institute of Astronomy and Astrophysics, Academia Sinica, 11F of Astronomy-Mathematics Building, No.1, Sec. 4, Roosevelt Road, Taipei 10617, Taiwan, R.O.C.}
\author[0000-0002-1765-7012]{Shusuke Onishi}
\affiliation{Institute of Space and Astronautical Science, Japan Aerospace Exploration Agency, 3-1-1 Yoshinodai, Chuo-ku, Sagamihara, Kanagawa 252-5210, Japan}
\author[0000-0002-6660-9375]{Takao Nakagawa}
\affiliation{Institute of Space and Astronautical Science, Japan Aerospace Exploration Agency, 3-1-1 Yoshinodai, Chuo-ku, Sagamihara, Kanagawa 252-5210, Japan}
\author[0000-0002-5012-6707]{Kosei Matsumoto}
\affiliation{Institute of Space and Astronautical Science, Japan Aerospace Exploration Agency, 3-1-1 Yoshinodai, Chuo-ku, Sagamihara, Kanagawa 252-5210, Japan}
\affiliation{Sterrenkundig Observatorium Department of Physics and Astronomy Universiteit Gent Krijgslaan 281 S9, B-9000 Gent, Belgium}
\affiliation{Department of Physics, Graduate School of Science, The University of Tokyo, 7-3-1 Hongo, Bunkyo-ku, Tokyo 113-0033, Japan}
\author{Naoki Isobe}
\affiliation{Institute of Space and Astronautical Science, Japan Aerospace Exploration Agency, 3-1-1 Yoshinodai, Chuo-ku, Sagamihara, Kanagawa 252-5210, Japan}
\author{Mai Shirahata}
\affiliation{Institute of Space and Astronautical Science, Japan Aerospace Exploration Agency, 3-1-1 Yoshinodai, Chuo-ku, Sagamihara, Kanagawa 252-5210, Japan}
\author[0000-0002-9850-6290]{Shunsuke Baba}
\affiliation{Kagoshima University, Graduate School of Science and Engineering, 1-21-35 Korimoto, Kagoshima, Kagoshima 890-0065, Japan}
\author[0000-0001-5187-2288]{Kazushi Sakamoto}
\affiliation{Institute of Astronomy and Astrophysics, Academia Sinica, 11F of Astronomy-Mathematics Building, No.1, Sec. 4, Roosevelt Rd, Taipei 10617, Taiwan, R.O.C.}

\begin{abstract}
We investigated the inner buried nucleus of a nearby luminous infrared galaxy NGC~4418 using high-resolution spectroscopy of fundamental carbon monoxide (CO) rovibrational absorptions around $4.67~\mu$m for the first time.
This method allowed us to examine the physical and kinematical properties in the hot inner region of this nucleus.
We detected a series of both very deep (partly saturated) $^{12}$CO and moderately deep (optically thin) $^{13}$CO absorption lines and inferred a large column density ($N_\mathrm{H2}=(5\pm3)\times10^{23}$~cm$^{-2}$ in front of the $5~\mu$m photosphere) of warm ($T_\mathrm{ex}\simeq170$~K) molecular gas by assuming an isothermal plane-parallel slab illuminated by a compact background MIR-emitting source.
We modeled that the warm CO absorber almost covers the central heating source and that it is an inner layer around the $5~\mu$m photosphere (at $r=$several~pc) of a compact shroud of gas and dust ($d\sim100$~pc).
The width of the absorption lines ($110$~km~s$^{-1}$) and their small deviation from the systemic velocity ($<10$~km~s$^{-1}$) are consistent with a warm and turbulent layer with little bulk motion in the radial direction.
\end{abstract}


\section{Introduction} \label{sec:introduction}

NGC~4418 is a nearby luminous infrared galaxy (LIRG) notable for its compact and luminous but highly obscured nucleus at a distance of $34$~Mpc ($V_\mathrm{sys}=2117$~km~s$^{-1}$ in the heliocentric system\footnote{This is based on optical stellar spectroscopy at the nucleus.
\cite{sakamoto+21b} found $V_\mathrm{sys}=2115$~km~s$^{-1}$ (after conversion to the heliocentric velocity in the optical definition) with the high-resolution ($1''$) (sub)millimeter molecular line measurement, in agreement (within $2$~km~s$^{-1}$) with the optical measurement.}; \citealt{ohyama+19};
note that at this distance, $1''$ corresponds to $165$~pc).
This galaxy hosts a compact bright mid-infrared (MIR) emitting core at the nucleus, which cannot be resolved at $\sim0\farcs3$ resolution \citep{evans+03,siebenmorgen+08,roche+15}.
The corresponding submillimeter and far-infrared (FIR) cores \citep{sakamoto+13,sakamoto+21,lutz+16} emit most of its large bolometric luminosity ($L_\mathrm{bol}\simeq1\times10^{11}~L_\mathrm{\sun}$; e.g., \citealt{ga+12}).
It displays a prominent red spectral energy distribution (SED) at MIR beyond $\sim5~\mu$m, but only a stellar SED at the shorter wavelengths (\citealt{spoon+01,evans+03,imanishi+04,siebenmorgen+08}).
This object is located at the end of the deepest $9.7~\mu$m absorption by amorphous silicate dust (hereafter, the $9.7~\mu$m absorption) and the smallest equivalent width of the polycyclic aromatic hydrocarbon (PAH) $6.2~\mu$m emission on the so-called Spoon diagram and is classified as 3A \citep{spoon+07}.
The very deep $9.7~\mu$m absorption and the prominent ice features at MIR are often discussed in the context of obscured heating sources such as buried young stellar objects (e.g., \citealt{dudley+97,spoon+01,spoon+04}).
All of these characteristics are very different from typical active galactic nuclei (AGNs) and starburst galaxies, and the nucleus of NGC~4418 is sometimes referred to as a compact obscured nucleus (CON; \citealt{costagliola+13,conquest2} and references therein).

Whether the CONs are powered by very compact young star clusters or AGNs is an open question, and the CON of NGC~4418 has been extensively studied in this regard due to its proximity and large FIR luminosity.
NGC~4418 has often been classified as an AGN on the basis of indirect observational evidence such as a compact radio core ($\lesssim1''$; e.g., \citealt{kewley+00}; but see below for the very long baseline interferometry, VLBI result), the extremely large MIR luminosity surface density \citep{evans+03,siebenmorgen+08}, and the very deep $9.7~\mu$m and prominent ice absorptions at MIR (e.g., \citealt{dudley+97,spoon+01}).
However, more rigorous searches for an obscured AGN have not yet provided clear answers.
In X-rays, NuSTAR found no hard X-ray emission, although an AGN still cannot be ruled out because an extremely large hydrogen column density ($N_\mathrm{H}>10^{25}$~cm$^{-2}$; \citealt{ga+12,costagliola+13,sakamoto+21}) can obscure even the hard X-rays below the current detection limit \citep{yamada+21}.
At (sub)millimeter wavelengths, a molecular emission line ratio of HCN/HCO$^+$ has been proposed as a powerful diagnosis of the central heating source (e.g., \citealt{kohno+01,imanishi+04}) on the basis of the XDR (X-ray dominating region) models (e.g., \citealt{meijerink+05}).
This technique has been applied to this galaxy nucleus, but the results are still controversial (e.g., \citealt{imanishi+04,aalto+07,imanishi+18}; see also \citealt{ga+12,costagliola+13}).
In radio, VLBI observations have resolved the radio core into multiple very compact blobs, clearly indicating that a single AGN alone cannot explain its nuclear activity \citep{varenius+14}.

The main goal of this study is to investigate the physical and kinematical properties of the molecular gas within the CON of NGC~4418.
We utilize a novel method to analyze the fundamental ($v=0\rightarrow1$) rovibrational absorption lines of carbon monoxide (CO) centered at $4.67~\mu$m with high-resolution spectroscopy.
\cite{imanishi+10} reported the deep CO absorption in NGC~4418 with the low-resolution AKARI spectrum.
This method has some advantages over the abovementioned methods.
Thanks to the compact hot dust distribution around the central heating source ($r\sim3.6$~pc or $0\farcs04$ FWHM; Section~\ref{sec:photosphere}), we can use a pencil beam to effectively eliminate contamination from, e.g., circum-nuclear star formation.
In addition, thanks to the many rovibrational transitions of CO within a small wavelength interval, we can analyze many transitions of the same molecular species at once, eliminating ambiguity regarding the abundance effect on the line ratio.
Finally, absorption line analysis with transitions from the vibrational ground state is much simpler than emission line analysis when complex excitation such as IR pumping is involved.

\cite{spoon+03} detected resolved rovibrational CO absorption lines in NGC~4945 and pioneered the analysis of the CO absorption to derive physical properties of the CO-absorbing gas in an extragalactic environment.
\cite{spoon+04} detected a blended broad CO absorption feature in the low-resolution Spitzer spectrum of IRAS~F00183$-$7111, an ultra-luminous infrared galaxy (ULIRG) hosting an AGN, and investigated the warm dense gas in the vicinity of the nucleus for the first time.
\cite{geballe+06} and \cite{shirahata+13} performed the similar CO absorption analysis to IRAS~08572+3915, another nearby ULIRG hosting an AGN, using high-resolution spectroscopy and performed a detailed study of the physical and kinematical conditions of the AGN dusty torus.
\cite{onishi+21} further studied this galaxy to examine the inflow and outflow of warm CO in detail, and \cite{matsumoto+22} used a theoretical model to demonstrate that such flow signatures can be observed.
\cite{baba+18} studied a sample of nearby obscured AGNs with low-resolution AKARI and Spitzer spectra to systematically examine the properties of the hot gas in AGNs (see also \citealt{baba+22}).
We will compare NGC~4418 with these galaxies in detail in Section~\ref{sec:comp_other_galaxies}.

\section{Observation and data reduction}

The M-band Echelle spectrum of NGC~4418 around $4.8~\mu$m was obtained with an IRCS spectrograph \citep{ircs1,ircs2} at the Subaru telescope on 2010 February 27 and 28 (UT).
We adjusted the grisms to include as many $^{12}$CO $P$ branch ($J\rightarrow J-1$) transitions as possible and a few $R$ branch ($J\rightarrow J+1$) transitions spanning over two wavelength coverages with a small wavelength overlap.
Combined with a $0\farcs54$-wide slit, this setup provided a spectral resolution of $R=5300$ (or $dV=57$~km~s$^{-1}$).
We took many short-exposure frames while dithering along the slit direction every 60 seconds, following a standard ``ABBA'' sequence.
Total exposure times were 68 and 76 minutes for the short- and long-wavelength coverage, respectively.
The standard star HR~5685 was observed every night for each spectral coverage.

Data reduction was performed in a standard way for Echelle spectral images obtained in the ABBA sequence.
Sky subtraction was performed by subtracting A--B and B--A pairs of the frames and the sky-subtracted frames were stacked after correcting for dithering offsets.
The extracted one-dimensional spectrum of the standard star for each night was used to calibrate the wavelength (by referencing narrow telluric absorption features) and to remove the telluric absorption features of the similarly extracted spectrum of NGC~4418 taken on the same night.
A small flux scaling was applied to stitch the two spectra together using the wavelength overlap between them (between $4.766~\mu$m and $4.780~\mu$m around $P(8)$ and $P(9)$ of $^{12}$CO; Section~\ref{sec:spectrum}).
The noise of the spectrum was estimated from the standard deviation of individual short-exposure frames.
The heliocentric velocity correction was applied to the wavelength and velocity of the final spectrum.

\section{The spectrum} \label{sec:spectrum}

\begin{figure*}
\centering
\epsscale{1.15}
\plotone{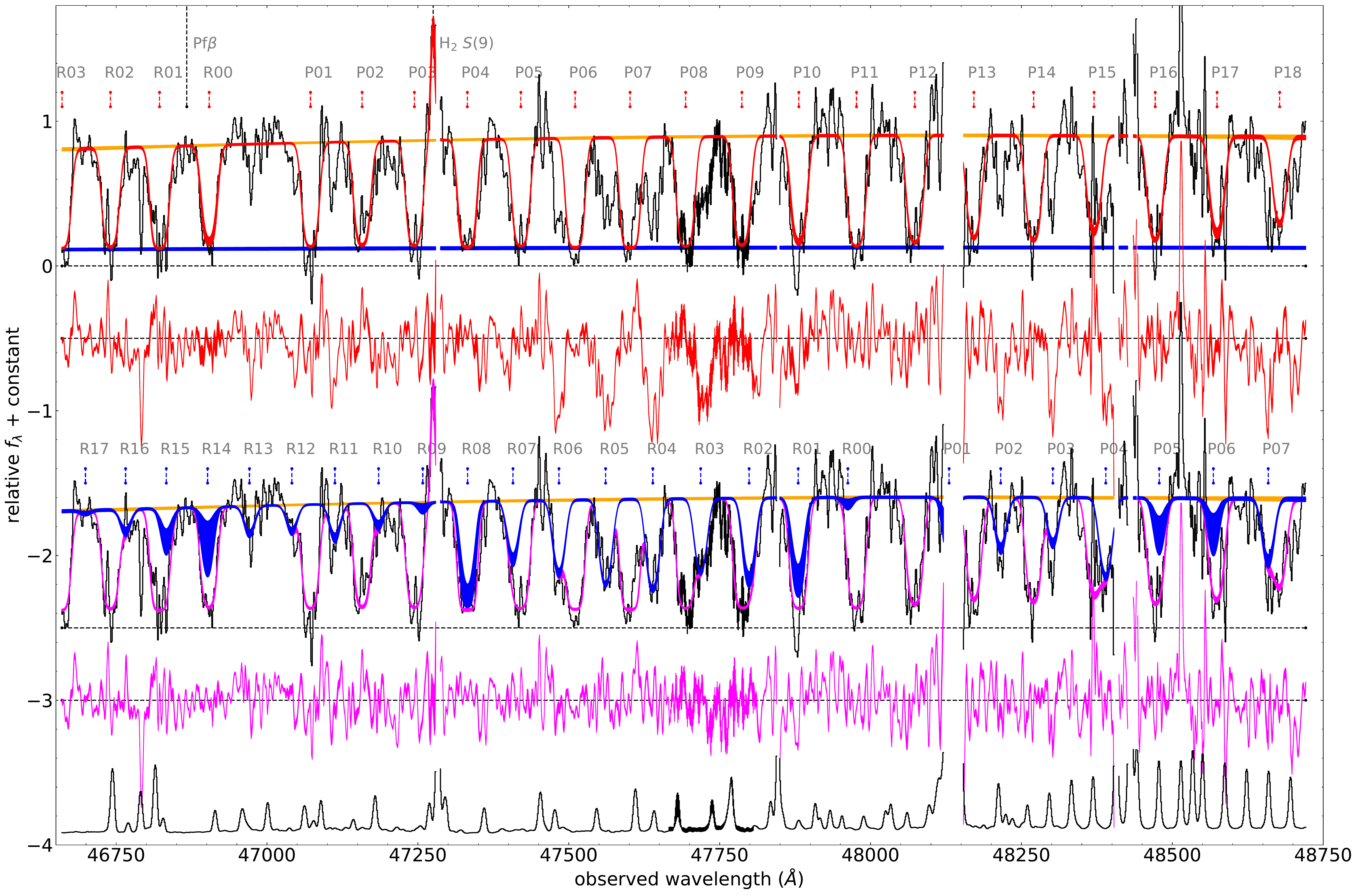}
\caption{
The observed and the best-fit model spectra with the full model.
Top: the observed spectrum (black) with the best-fit model spectra without $^{13}$CO (red), the continuum only (orange), and the floor only (blue) overlaid.
The detected $^{12}$CO and H$_2\,S(9)$ lines and the expected position of Pf$\beta$ for the CO velocity are marked.
Second: the residual spectrum with the best-fit model spectrum without $^{13}$CO (black$-$red in the top panel; red).
Third: the observed spectrum (black) with the best-fit model spectra including all components (magenta), without $^{12}$CO (blue), and the continuum only (orange) overlaid.
The modeled $^{13}$CO absorptions are marked.
Fourth: the residual spectrum including all components (black$-$magenta in the third panel; magenta).
Bottom: the noise spectrum (black).
The second, third, fourth, and bottom panels are shifted by $-0.5$, $-2.5$, $-3.0$, and $-4.0$, respectively, as indicated by the horizontal dashed lines.
All spectra are shown only where the signal-to-noise ratio (S/N) per pixel of the observed spectrum is greater than $1.5$, while the entire spectrum along with the noise was used for the model fit.
An emission line-like feature at $4.8515~\mu$m (or at $4.8173~\mu$m in the rest frame of NGC~4418) is located at one of the strong telluric absorption features and is an artifact caused by the telluric correction.
Each model spectrum has a shade corresponding to the $16$th--$84$th percentile range, which is very narrow except around some $^{13}$CO absorptions that overlap the nearby $^{12}$CO absorptions.
}
\label{fig:fig1}
\end{figure*}

The spectrum displays a series of deep absorption lines of $^{12}$CO $v=0\rightarrow1$ rovibrational transitions (hereafter, the $^{12}$CO absorptions; Figure~\ref{fig:fig1} top and third panels).
The noise spectrum (bottom panel) exhibits many narrow wavelength regions with higher noise, which are caused by strong telluric absorptions.
Their width corresponds to the instrumental resolution.
The $^{12}$CO absorption lines are much broader than these noise spikes and are found in a different pattern of the noise spikes, thereby confirming the robust detection of all of the $^{12}$CO absorptions.
Most of the narrow positive and negative spikes in the spectrum (top and third panels) that have similar widths to the noise spikes, on the other hand, are likely artifacts due to these strong telluric features.
Our wavelength coverage includes the $P$ branch up to $P(18)$ and the $R$ branch up to $R(3)$ of $^{12}$CO.
Many $^{12}$CO absorptions are deep, but display nonzero flux at the bottom of the absorption profiles.
The spectrum around $^{12}$CO $P(6)$--$P(8)$ (and some other wavelength regions) looks more complex than other spectral regions that simply display periodic $^{12}$CO absorptions, indicating additional absorption features there.
They appear in a different pattern of the noise spikes, as in the case of the $^{12}$CO absorptions.
As we will demonstrate below, they are explained by the $^{13}$CO $v=0\rightarrow1$ rovibrational absorption lines (hereafter, the $^{13}$CO absorptions).
Our wavelength coverage includes the $P$ branch up to $P(7)$ and the $R$ branch up to $R(17)$ of $^{13}$CO.
The $^{13}$CO absorptions, as some isolated ones (e.g., $P(2)$ and $R(5)$) indicate, are much shallower than the $^{12}$CO absorptions.
We also detected a pure-rotational H$_2\,S(9)$ emission line, but no hydrogen Pf$\beta$ emission line.

\subsection{Spectral fitting} \label{sec:spectral_fitting}

We performed spectral model fitting to measure the optical depths, the recession velocity, and the line width of the $^{12}$CO and $^{13}$CO absorptions.
We adopted a realistic but as simple as possible model in this work, while S. Onishi et al. (2023, in preparation) will discuss the multicomponent models.
We assumed a single velocity component, but fitted the $^{12}$CO and $^{13}$CO absorptions independently to measure $\tau_\mathrm{12CO}(J)$ and $\tau_\mathrm{13CO}(J)$.
We also assumed a Gaussian optical depth profile for each of the CO absorption lines with the common recession velocity ($V_\mathrm{CO}$) and the velocity width ($\sigma_\mathrm{CO}$) for the $^{12}$CO and $^{13}$CO absorptions.
We obtained their transition parameters from HITRAN \citep{hitran} and the Leiden Atomic and Molecular Database \citep{LAMBDA}.
We fitted the spectrum with a second-order polynomial continuum without the continuum normalization.
We assumed an isothermal plane-parallel slab illuminated by a compact background MIR-emitting source.
We modeled the observed spectrum by the background continuum with the CO absorption lines and a scaled continuum, as
$f_\mathrm{obs}\equiv(1-S)\times f_\mathrm{cont}\times\exp{(-\sum(\tau_\mathrm{CO(J)})})+S\times f_\mathrm{cont}$,
where $S$ represents the scaling constant.
The scaled continuum in the second term (the continuum ``floor'') is to approximate the nonzero fluxes at the bottom of the deep $^{12}$CO absorptions.
We caution that, in general, both CO line emission and dust continuum emission from various sources contribute to the floor\footnote{
For the lines, CO emission from the same CO-absorbing region by absorbing dust continuum from directions other than the line of sight of the observer, stimulated CO emission from the same CO-absorbing region by absorbing the background light, and CO emission from the nearby (inside the slit but outside the background source) molecular gas.
As for the continuum, the continuum emission from colder dust in front of the CO-absorbing region, and the unabsorbed background continuum emission when the absorber covers only a part of the background light source.
}, and we cannot distinguish them without detailed physical models of gas, dust, and their emission.
The floor can be simplified as the leaking background continuum not covered by the absorber in front, as
$(1-C_\mathrm{f})\times f_\mathrm{cont}$,
where $C_\mathrm{f}$ is the covering factor of the absorber over the background light when seen from the observer, if we neglect all other contributions.
In the following, we adopt $S=1-C_\mathrm{f}$ for simplicity.
We note that $C_\mathrm{f}$ measured this way is a lower limit of the true covering factor in general, and the less-than-unity $C_\mathrm{f}$ we obtain below does not necessarily indicate the presence of uncovered light from the background source.
We added the H$_2\,S(9)$ emission line by assuming a Gaussian profile at a recession velocity ($V_\mathrm{H2S9}$) different from $V_\mathrm{CO}$.
Because this line appears as narrow as the instrumental width (see below), we assumed it to be unresolved.
Possible ice absorptions within the wavelength coverage, two CO ice features near $^{12}$CO $P(1)$, are much broader than the CO gas absorption lines (see \citealt{onishi+21} for the case of IRAS~08572+3915), and the spectrum does not display such features.
Therefore, we omitted them from the model.
We call the model described so far the full model to distinguish it from the simplified model described below.

We simultaneously fitted the continuum, the floor, and all spectral features with 55 free parameters with the full model.
We utilized the Bayesian fitting code ``emcee'', which incorporates the Markov Chain Monte Carlo technique \citep{emcee}.
We adopted flat priors with the following simple range constraints for physically likely solutions and better convergence.\footnote{We also applied the following trivial constraints:
$\tau\ge0$ and $\sigma\ge0$ for all absorption lines, amplitude$\ge0$ and $\sigma\ge0$ for the H$_2\,S(9)$ emission line, and $0\le C_\mathrm{f}\le1$.}
For some $^{12}$CO and $^{13}$CO pairs whose absorption profiles closely overlap each other, we set $\tau_\mathrm{12CO}>\tau_\mathrm{13CO}$.
We adopted a criterion of close overlap as the central wavelength differences of a line pair being less than $15$~km~s$^{-1}$ ($\simeq1/4\times\sigma_\mathrm{CO}$).
For H$_2\,S(9)$, we constrained its recession velocity to avoid the nearby CO absorption lines.
In the emcee run, we had $120$ walkers, each with $5\times10^5$ steps, thinned every $500$ steps, with the burn-in steps removed by $1\times10^5$, for a typical autocorrelation length of $2000$.

We also performed a similar spectral model fitting, but with a simplified model assuming a single component in the local thermodynamic equilibrium (LTE) conditions (hereafter, the LTE model), to test the robustness of the fitting with the full model above.
This is because the full model, with a very large number of free parameters to fit (55), could overfit the complicated spectrum and/or be severely affected by degeneracy among the fitting parameters, and could provide unreliable results, although the full model results indicate a simple LTE-like population distribution on the rotation diagrams (Section~\ref{sec:rotation_diagram}).
In the LTE model, we linked the optical depths of different transitions according to the Boltzmann distribution and derived the excitation temperatures and the column densities directly, without fitting the optical depths on the rotation diagrams afterward.
We adopted the same assumptions and model equation for the kinematics and the line profile of the CO absorptions, the continuum shape, the floor, and the H$_2$ emission line as for the full model.
This simplified model requires only 12 parameters to fit (six parameters regarding the CO absorption lines---one set of column density and excitation temperature for $^{12}$CO, another set for $^{13}$CO, their common recession velocity and the velocity width---along with six parameters regarding other components---the continuum, the floor, and the H$_2$ emission---that are common to the full model), ensuring much more robust results.
We used the same software in a similar manner as for the full model.
The LTE model results and comparison with the full model results are shown in the Appendix.
Because we found consistent results between the two models, we adopt the full model results in the following sections.

\subsection{Fit results} \label{sec:spectral_fit_result}

With the full model, we successfully obtained a good fit with reduced-$\chi^2\simeq1.1$ with 1985 degrees of freedom (Figure~\ref{fig:fig1}; Table~\ref{tab:tab1}).
Most parameters, except those with the closely overlapping $^{12}$CO and $^{13}$CO absorptions, were well constrained.
The best-fit model reasonably reproduced some spectral portions between the CO absorption lines that display an unabsorbed continuum (e.g., $R(0)$--$P(1)$ and $P(10)$--$P(12)$ of $^{12}$CO).
This suggests that the true continuum is not far above the best-fit model continuum and that the CO absorption depths were not significantly underestimated due to the inaccurate continuum placement.
The $^{12}$CO absorptions are deep and mostly saturated (over a floor with a large $C_\mathrm{f}=0.86\pm0.01$) with $\tau\gtrsim3$ at the line centers (top panel).
Even higher-$J$ absorption lines (up to $P(18)$ with $E_\mathrm{lower}(J)/k=945$~K, where $k$ is the Boltzmann constant) are deep although they may not be as heavily saturated as the lower-$J$ ones.
The residual spectrum after subtracting the $^{12}$CO absorptions (second panel) reveals a series of moderately deep ($\tau\lesssim1$ at the line centers) absorption lines whose wavelengths exactly match those of the $^{13}$CO absorptions (third and fourth panels).
After removing uncertain detections due to deep nearby $^{12}$CO absorptions or low signal-to-noise ratio (S/N), the $^{13}$CO absorptions were unambiguously detected up to $J=13$ ($R(13)$) with $E_\mathrm{lower}(J)/k=481$~K.
The recession velocity for all CO absorptions is very close to the systemic one ($dV_\mathrm{CO}=+4\pm1$~km~s$^{-1}$).
The absorption profiles are broad ($110\pm3$~km~s$^{-1}$ FWHM after subtracting the instrumental width in quadrature), as some isolated unsaturated $^{13}$CO absorptions (e.g., $P(2)$ and $R(5)$) indicate.
The H$_2\,S(9)$ emission line is well reproduced by an unresolved line (FWHM$=57$~km~s$^{-1}$) at a slightly lower velocity than the systemic one ($dV_\mathrm{H2S9}=-21\pm7$~km~s$^{-1}$).

\startlongtable
\begin{deluxetable*}{ccccccccccc}
\tablewidth{0pt} 
\tablecaption{Parameters of the CO Absorption Lines ($v=0\rightarrow1$, $J\rightarrow J'$) toward NGC~4418\label{tab:tab1}}
\tablehead{
\colhead{$J$} & \colhead{$\lambda$\tablenotemark{a}} & \colhead{$E_\mathrm{lower}(J)/k$\tablenotemark{b}} & \colhead{$F_\mathrm{JJ'}$\tablenotemark{c}} & \multicolumn{3}{c}{$\tau$\tablenotemark{d}} & \multicolumn{3}{c}{$N_\mathrm{CO}(J)$\tablenotemark{e} ($10^{17}$~cm$^{-2}$)} & \colhead{Rotation Diagram\tablenotemark{f}}\\
\colhead{} & \colhead{(\AA)} & \colhead{(K)} & \colhead{($10^{-6})$} & \colhead{16th} & \colhead{50th} & \colhead{84th} & \colhead{16th} & \colhead{50th} & \colhead{84th} & \colhead{} \\
\colhead{} & \colhead{     } & \colhead{   } & \colhead{           } & \colhead{Percentile} & \colhead{Percentile} & \colhead{Percentile} & \colhead{Percentile} & \colhead{Percentile} & \colhead{Percentile} & \colhead{}
}
\startdata
\multicolumn{10}{l}{$^{12}$CO $R$ branch ($J'=J+1$)}\\
3 & 46333 & 33.1917 & 6.5668 & 82.9 & 92.8 & 104.4 & 6.64 & 7.44 & 8.36 & \checkmark \\
2 & 46412 & 16.5962 & 6.8814 & 70.7 & 80.6 & 92.3 & 5.39 & 6.14 & 7.03 & \checkmark \\
1 & 46493 & 5.5321 & 7.6270 & 95.8 & 115.9 & 139.5 & 6.57 & 7.94 & 9.56 \\
0 & 46575 & 0.0000 & 11.4173 & 44.5 & 57.8 & 71.2 & 2.03 & 2.64 & 3.25 \\
\hline
\multicolumn{10}{l}{$^{12}$CO $P$ branch ($J'=J-1$)}\\
1 & 46742 & 5.5321 & 3.7916 & 75.7 & 86.4 & 99.3 & 10.32 & 11.78 & 13.54 & \checkmark \\
2 & 46826 & 16.5962 & 4.5415 & 63.3 & 74.8 & 90.6 & 7.18 & 8.49 & 10.28 & \checkmark \\
3 & 46912 & 33.1917 & 4.8565 & 72.6 & 80.9 & 90.4 & 7.68 & 8.55 & 9.55 \\
4 & 46999 & 55.3180 & 5.0246 & 90.6 & 114.2 & 149.1 & 9.22 & 11.62 & 15.18 \\
5 & 47088 & 82.9744 & 5.1288 & 78.3 & 89.5 & 103.1 & 7.78 & 8.89 & 10.24 \\
6 & 47177 & 116.1597 & 5.1973 & 108.7 & 123.4 & 141.4 & 10.61 & 12.05 & 13.81 & \checkmark \\
7 & 47267 & 154.8727 & 5.2451 & 136.2 & 161.6 & 194.1 & 13.13 & 15.58 & 18.71 & \checkmark \\
8 & 47359 & 199.1120 & 5.2807 & 85.5 & 97.5 & 111.8 & 8.16 & 9.30 & 10.66 & \checkmark \\
9 & 47451 & 248.8756 & 5.3033 & 70.1 & 80.9 & 94.8 & 6.63 & 7.65 & 8.96 \\
10 & 47545 & 304.1619 & 5.3196 & 52.8 & 63.7 & 81.2 & 4.96 & 5.99 & 7.63 \\
11 & 47640 & 364.9688 & 5.3307 & 85.9 & 97.5 & 111.7 & 8.02 & 9.10 & 10.43 \\
12 & 47736 & 431.2938 & 5.3396 & 60.5 & 68.6 & 78.5 & 5.62 & 6.37 & 7.29 & \checkmark \\
13 & 47833 & 503.1343 & 5.3483 & 47.3 & 52.0 & 57.5 & 4.36 & 4.80 & 5.31 & \checkmark \\
14 & 47931 & 580.4879 & 5.3486 & 53.6 & 60.0 & 67.5 & 4.93 & 5.51 & 6.21 & \checkmark \\
15 & 48031 & 663.3513 & 5.3512 & 35.8 & 41.7 & 48.8 & 3.28 & 3.82 & 4.46 & \checkmark \\
16 & 48131 & 751.7215 & 5.3505 & 50.2 & 58.4 & 67.9 & 4.58 & 5.33 & 6.19 \\
17 & 48233 & 845.5952 & 5.3502 & 35.4 & 45.2 & 55.1 & 3.22 & 4.10 & 5.00 \\
18 & 48336 & 944.9686 & 5.3477 & 29.7 & 33.1 & 37.1 & 2.68 & 3.00 & 3.35 & \checkmark \\
\hline
\multicolumn{10}{l}{$^{13}$CO $R$ branch ($J'=J+1$)}\\
17 & 46371 & 808.4389 & 5.7966 & 0.2 & 0.6 & 1.2 & 0.01 & 0.05 & 0.11 & \checkmark \\
16 & 46437 & 718.6866 & 5.7945 & 3.9 & 5.2 & 6.6 & 0.35 & 0.47 & 0.60 & \checkmark \\
15 & 46504 & 634.1972 & 5.7981 & 4.9 & 8.5 & 12.1 & 0.45 & 0.76 & 1.09 \\
14 & 46572 & 554.9732 & 5.8012 & 5.3 & 14.7 & 25.4 & 0.48 & 1.32 & 2.28 \\
13 & 46641 & 481.0177 & 5.8006 & 5.3 & 6.6 & 8.0 & 0.48 & 0.59 & 0.71 & \checkmark \\
12 & 46711 & 412.3335 & 5.8078 & 4.8 & 5.9 & 7.0 & 0.43 & 0.52 & 0.62 & \checkmark \\
11 & 46782 & 348.9230 & 5.8201 & 6.8 & 7.9 & 9.2 & 0.60 & 0.70 & 0.81 & \checkmark \\
10 & 46853 & 290.7887 & 5.8285 & 3.6 & 5.0 & 6.6 & 0.32 & 0.45 & 0.58 & \checkmark \\
9 & 46926 & 237.9326 & 5.8454 & 0.3 & 1.0 & 2.4 & 0.02 & 0.09 & 0.21 \\
8 & 47000 & 190.3565 & 5.8703 & 31.3 & 61.7 & 83.4 & 2.73 & 5.38 & 7.27 \\
7 & 47075 & 148.0622 & 5.9039 & 13.2 & 16.1 & 19.4 & 1.14 & 1.39 & 1.67 \\
6 & 47150 & 111.0515 & 5.9455 & 19.8 & 22.5 & 25.5 & 1.70 & 1.92 & 2.18 & \checkmark \\
5 & 47227 & 79.3253 & 6.0147 & 28.3 & 31.0 & 33.8 & 2.39 & 2.61 & 2.85 & \checkmark \\
4 & 47305 & 52.8852 & 6.1138 & 32.7 & 35.9 & 39.4 & 2.70 & 2.96 & 3.25 & \checkmark \\
3 & 47383 & 31.7319 & 6.2755 & 21.5 & 23.2 & 25.1 & 1.72 & 1.86 & 2.01 & \checkmark \\
2 & 47463 & 15.8662 & 6.5775 & 22.9 & 27.6 & 33.0 & 1.75 & 2.10 & 2.51 \\
1 & 47544 & 5.2888 & 7.2983 & 21.0 & 37.4 & 47.6 & 1.44 & 2.56 & 3.26 \\
0 & 47626 & 0.0000 & 10.9188 & 0.3 & 1.0 & 2.3 & 0.01 & 0.04 & 0.11 \\
\hline
\multicolumn{10}{l}{$^{13}$CO $P$ branch ($J'=J-1$)}\\
1 & 47792 & 5.2888 & 3.6252 & 2.7 & 9.5 & 21.7 & 0.37 & 1.29 & 2.96 \\
2 & 47877 & 15.8662 & 4.3416 & 10.6 & 12.8 & 15.2 & 1.21 & 1.45 & 1.72 & \checkmark \\
3 & 47963 & 31.7319 & 4.6431 & 9.7 & 11.4 & 13.1 & 1.02 & 1.20 & 1.39 & \checkmark \\
4 & 48050 & 52.8852 & 4.8078 & 23.1 & 26.3 & 29.8 & 2.36 & 2.67 & 3.03 & \checkmark \\
5 & 48138 & 79.3253 & 4.9057 & 3.9 & 8.6 & 14.0 & 0.38 & 0.85 & 1.39 \\
6 & 48227 & 111.0515 & 4.9713 & 2.9 & 8.4 & 15.3 & 0.29 & 0.82 & 1.50 \\
7 & 48317 & 148.0622 & 5.0169 & 14.5 & 17.5 & 20.9 & 1.40 & 1.69 & 2.01 & \checkmark
\enddata

\tablenotetext{a}{Rest-frame wavelength.}
\tablenotetext{b}{Lower-state energy in temperature.}
\tablenotetext{c}{Oscillator strength.}
\tablenotetext{d}{Optical depths.}
\tablenotetext{e}{Column densities.}
\tablenotetext{f}{The absorption lines used for the rotation diagram analysis (see the main text) are checked.}

\end{deluxetable*}

\section{Rotation Diagrams and Hydrogen Column Density}\label{sec:rotation_diagram}

\begin{figure}
\centering
\epsscale{1.15}
\plotone{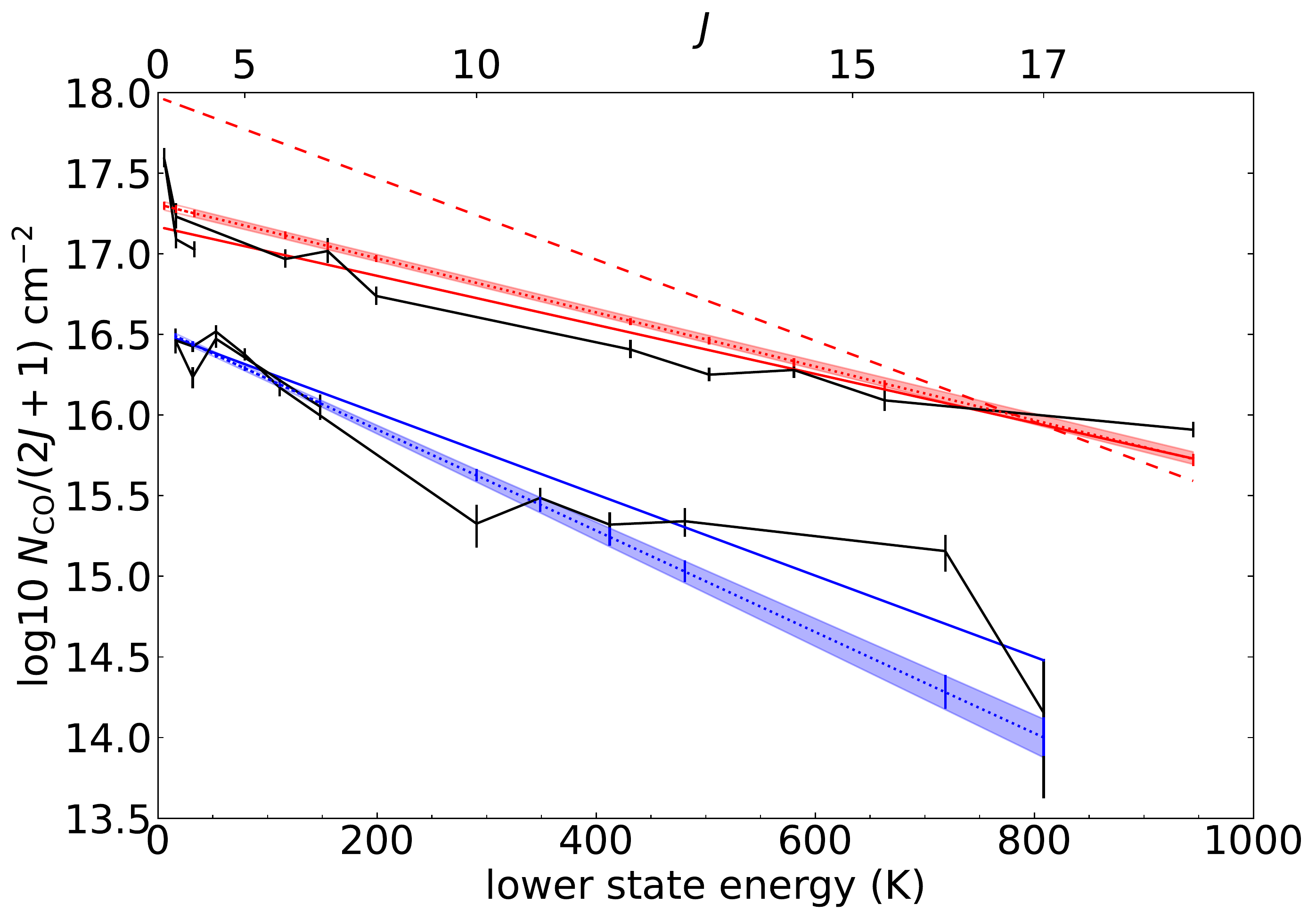}
\caption{
The rotation diagrams of $^{12}$CO (top) and $^{13}$CO (bottom) in log10 scale with $16$th--$84$th percentile ranges around the median as a function of $E_\mathrm{lower}(J)/k$.
The red and blue solid straight lines represent the best-fit LTE populations with the full model result for the $^{12}$CO and $^{13}$CO, respectively.
The red dashed line represents the scaled (by a factor of $29$) $^{13}$CO best-fit model (blue solid line).
The red and blue dotted lines with shaded areas represent the best-fit populations with the LTE model (see the Appendix) for $^{12}$CO and $^{13}$CO, respectively.
The shaded areas represent the 16th--84th percentile ranges of the populations.
}
\label{fig:fig2}
\end{figure}

We analyzed the rotational-level populations in the $v=0$ vibrational energy state on the rotation diagrams (Figure~\ref{fig:fig2}).
We calculated the column density of the $J$th rotational state ($N_\mathrm{CO}(J)$) using the optical depth of that transition with equations (2)--(5) of \cite{onishi+21}.
We removed the optical depth measurements of $^{12}$CO $P(3)$ and $^{13}$CO $R(9)$ immediately adjacent to the H$_2\,S(9)$ emission, as well as some closely overlapping $^{12}$CO and $^{13}$CO pairs (Section~\ref{sec:spectral_fitting}; see the last column of Table~\ref{tab:tab1} for the transitions used).
When the gas is in the LTE conditions, the level populations are distributed according to the Boltzmann distribution, as in Equation (12) of \cite{onishi+21}.
Under such conditions, the (natural) logarithm of level populations (divided by the statistical weight of $2J+1$) align along a straight line determined by the excitation temperature ($T_\mathrm{ex}$) and the total column.

We found that linear functions fit both $^{12}$CO and $^{13}$CO reasonably well and derived $T_\mathrm{ex,12CO}=286\pm30$~K and $T_\mathrm{ex,13CO}=173\pm17$~K, and the total column densities $N_\mathrm{12CO}=(1.5\pm0.3)\times10^{19}$~cm$^{-2}$ and $N_\mathrm{13CO}=(4.3\pm0.5)\times10^{18}$~cm$^{-2}$.
However, the results seem unreasonable;
the two excitation temperatures do not match, and the $^{12}$CO column density is only $\simeq3.5$ times that of $^{13}$CO, while the Galactic $[^{12}\mathrm{C}]/[^{13}\mathrm{C}]$ abundance ratio is $20$--$100$ (e.g., \citealt{yan+23}).
We interpreted the results as an underestimation of the $^{12}$CO columns of the lower-$J$ levels due to the saturation of the absorptions;
the deep absorption profiles change little as a function of the optical depth near the bottoms, where the S/N is very low due to the absorptions, and the fit cannot exclude solutions with moderate optical depths.
The posterior probability distribution functions of the lower-$J$ $^{12}$CO columns indeed show asymmetric tails toward larger columns, while those of the $^{13}$CO columns do not.
Therefore, we scaled the $^{13}$CO model-level populations to match the three highest-$J$ $^{12}$CO columns by assuming $T_\mathrm{ex,12CO}=T_\mathrm{ex,13CO}$ and obtained $N_\mathrm{12CO}/N_\mathrm{13CO}=29\pm15$ and $N_\mathrm{12CO}\sim(6\pm3)\times10^{19}$~cm$^{-2}$.
Adopting $\mathrm{log}(N_\mathrm{H2}/N_\mathrm{CO})=3.9$, a typical value in the photon dominated region (PDR) and XDR models of \cite{meijerink+05} for large hydrogen column densities, we obtained the H$_2$ column density $N_\mathrm{H2}=(5\pm3)\times10^{23}$~cm$^{-2}$ for the warm CO absorber.
This column density is only up to the $5~\mu$m photosphere that emits the background light (Section~\ref{sec:photosphere}; see also Section~\ref{sec:where_CO}).

The largest systematic uncertainty in deriving the hydrogen column density is in the $[^{12}\mathrm{CO}]/[^{13}\mathrm{CO}]$ abundance ratio.
This ratio varies by more than a factor of 5 within our Galaxy (e.g., \citealt{yan+23}), and our adopted ratio ($N_\mathrm{12CO}/N_\mathrm{13CO}=29\pm15$) is close to the carbon isotope ratio of $[^{12}\mathrm{C}]/[^{13}\mathrm{C}]=21\pm5$ near our Galactic center \citep{yan+23}.
The lower abundance ratio there is attributed to more star formation in the past than in the outer Galactic radius.
Because the nucleus of NGC~4418 has likely experienced active star formation in the recent past \citep{ohyama+19}, our column estimate seems reasonable.

\section{Comparison with the AKARI Low-resolution Spectrum}

Our high-resolution measurement of the CO absorption lines is consistent with the low-resolution ($R\simeq160$ at $4.7~\mu$m) AKARI CO measurement of the blended CO absorption feature.
\cite{imanishi+10} reported the AKARI CO absorption depth $\tau(\mathrm{CO})=0.5$.
Here, $\tau(\mathrm{CO})$ is the apparent depth of the blended CO feature measured at the bottom of the blended $R$-branch feature\footnote{
Many lines in the $P$- and $R$-branches are blended in the low-resolution AKARI spectra to form a broad two-horned absorption feature centered at $\simeq4.7~\mu$m.
The blended $R$-branch feature is generally deeper than the blended $P$-branch feature due to larger oscillator strengths and smaller wavelength intervals between the individual absorption lines in the $R$-branch (Table~\ref{tab:tab1}).},
and is often used as an indicator of the absorbing CO column density without careful analysis using LTE spectral modeling \citep{spoon+04,baba+18,baba+22}.
We note, however, that $\tau(\mathrm{CO})$ is proportional to the CO column density only until individual CO absorption lines are saturated; once saturated, it increases with the line (velocity) width (see \citealt{baba+18} for the demonstration).
We adopted our best-fit LTE model (see the Appendix) to simulate the AKARI spectrum; we first synthesized the full CO absorption spectrum including the $J>3$ $R$-branch transitions below our wavelength coverage and then applied spectral smoothing.
We obtained $\tau(\mathrm{CO})=0.5$, in good agreement with the actual AKARI CO measurement \citep{imanishi+10}.

\section{Discussion}

\subsection{Physical Conditions of the CO Gas} \label{sec:physical_conditions}

The level populations on the population diagram are roughly distributed along a straight line (Section~\ref{sec:rotation_diagram}), suggesting the possibility of the LTE conditions where only collisions determine the level populations.
\cite{mashian+15} studied the high-$J$ ($14$--$19$) level populations using the CO spectral line energy distribution (SLED) with the Herschel data and found that a simple large velocity gradient model with warm ($T_\mathrm{kin}=63$--$125$~K) and dense ($n_\mathrm{H2}=4\times10^{5}$~cm$^{-3}$) gas, with no background radiation field, can explain the SLED.
A recent high-resolution Atacama Large Millimeter/submillimeter Array (ALMA) study revealed a sharp CO concentration in ($J=2$--$1$) -- ($J=6$--$5$) rotational transitions at the nucleus (with $0\farcs15$--$0\farcs3$ beams; \citealt{sakamoto+21b}), strongly suggesting that the warm dense gas is associated with the nuclear region (see also Section~\ref{sec:where_CO}).
\cite{ga+12} estimated the gas density $n_\mathrm{H2}=3\times10^6$~cm$^{-3}$ in their ``core'' region ($T_\mathrm{d}=140$--$150$~K, $r=10$~pc)\footnote{
This region surrounds the innermost ``hot'' region ($T_\mathrm{d}=350$~K, $r=2.6$~pc), which predominantly emits the MIR SED.
See Section~\ref{sec:photosphere}.}
on the basis of the detailed spectral analysis of multiple FIR and submillimeter lines.
For comparison, the highest $^{13}$CO rotational level we firmly detected in rovibrational absorption is $J=13$ (in $R(13)$), and the critical density for the collisional excitation to this level is $n_\mathrm{cr}(J=13)\sim2\times10^6$~cm$^{-3}$.\footnote{
We adopted the two-level approximation $n_\mathrm{cr}(J)=A_\mathrm{JJ'}/\gamma_\mathrm{JJ'}$, where $A_\mathrm{JJ'}$ is Einstein's A coefficient and $\gamma_\mathrm{JJ'}$ is a collision rate for the $J\rightarrow J'$ transition \citep{shirley15}.}
This is of the same order as the density measured by \cite{ga+12}.
Therefore, the CO excitation is likely dominated by collisional excitation, and the CO molecules there are warm ($T_\mathrm{kin}\simeq T_\mathrm{ex}$).
We note that despite the ambiguity of the CO gas temperature, the measured columns are roughly correct unless the higher-$J$ absorption lines beyond our spectral coverage are much stronger or weaker than simple extrapolation from the current coverage.
S. Onishi et al. (2023, in preparation) will present a more detailed analysis of the excitation.

\subsection{$5~\mu$m Dust Photosphere}\label{sec:photosphere}

Most of the background $5~\mu$m emission originates from a $5~\mu$m dust photosphere because the hotter interior cannot be seen behind an optically thick dust shroud with a unity covering factor around the nucleus (Sections~\ref{sec:where_CO}, \ref{sec:buried_N4418}).
\cite{dudley+97} measured the temperature and the effective size of the photosphere for the $\sim10~\mu$m light ($T_\mathrm{d}=280$~K, $r=0.6$~pc) using their continuum flux measurements at $8~\mu$m and $14~\mu$m;
observers see the same photosphere (with a dust opacity of $\simeq1$) at these two wavelengths on either side of the opacity bump around $9.7~\mu$m.
\cite{ga+12} estimated the temperature and the size of their innermost ``hot'' region ($T_\mathrm{d}=350$~K, $r=2.6$~pc) and the foreground absorption ($A_\mathrm{V}=70$ mag) to reproduce the Spitzer MIR SED.
Utilizing the better Spitzer SED of \cite{ga+12} and employing the more robust method of \cite{dudley+97} that is insensitive to the absorption modeling and fitting, we obtained the updated parameters ($T_\mathrm{d}=300$~K, $r=3.6$~pc assuming $\tau_\mathrm{8\mu m}=\tau_\mathrm{14\mu m}=1$).
For comparison, according to a simple model of spherical dust enshrouding a central heating source with $L_\mathrm{bol}=10^{11}~L_\mathrm{\sun}$ (\citealt{scoville13}), $T_\mathrm{d}=200$--$350$~K at $r=3.6$~pc for the optically thin and thick limits, respectively.\footnote{
The dust is heated by the central heating source and the neighboring heated dust in the optically thin and thick limits, respectively.}
Therefore, the $10~\mu$m background light is likely from an optically thick dust sphere with $T_\mathrm{d}\simeq300$~K at $r\sim4$~pc.
We note that the $N$-band ($\sim10~\mu$m) emission is pointlike with a $\simeq0\farcs2$ ($r\simeq16$~pc) FWHM beam \citep{evans+03,siebenmorgen+08,roche+15}, in agreement with this picture.
The dust emission has a nuclear core of FWHM$\sim60$~mas ($10$~pc) and peak $T_\mathrm{b}\sim400$~K at $440$ and $860~\micron$, where one penetrates the nucleus deeper than in MIR \citep{sakamoto+21}.
For our qualitative discussion below, we assume that the size of the $5~\mu$m photosphere is approximately the same as that of the $10~\mu$m photosphere ($r_\mathrm{5\mu m}\sim r_\mathrm{10\mu m}\sim4$~pc).
This size defines our $\sim0\farcs04$ beam (for $r\sim3.6$~pc) to sample CO in absorption.

\subsection{Where Is the Warm CO-absorbing Region?}\label{sec:where_CO}

The distribution of the warm CO-absorbing gas can be constrained by the high-resolution CO emission line study with ALMA \citep{sakamoto+21b}.
The CO ($J=3$--$2$) peak brightness temperature of the nuclear molecular-gas concentration measured with a $0\farcs15$ FWHM ($r_\mathrm{0\farcs15}=12$~pc) beam is $T_\mathrm{b}=144$~K (without the correction for the possible beam dilution and with nominal continuum subtraction; the true peak $T_\mathrm{b}$ is likely higher for these reasons; \citealt{sakamoto+21b}).
Although this brightness temperature is lower than the excitation temperature we measured ($T_\mathrm{ex}=173\pm17$~K), the warm CO-absorbing gas must be located within this nuclear concentration, because no other circum-nuclear regions display $T_\mathrm{b}>100$~K.
If the CO excitation condition is close to LTE, the similarity between the dust photosphere temperature and the CO excitation and brightness temperatures suggests that the CO-absorbing layer is close and outside the photosphere ($r=$several parsecs): this causes CO to be absorbed.

The warm CO-absorbing gas is most likely located between $r_\mathrm{5\mu m}$ and $r_\mathrm{0\farcs15}$ ($4$--$12$~pc) and, thus, within a compact dust shroud around the central heating source.
The dust shrouds of NGC~4418 and some other sources in the same class 3A (NGC~1377 and IRAS~08572+3915; \citealt{spoon+07}) have been modeled with a compact (with the inner radius $r_\mathrm{in}$ at the dust sublimation radius $r_\mathrm{sub}$) but geometrically thick ($r_\mathrm{out}/r_\mathrm{in}\ge100$, where $r_\mathrm{out}$ is the outer radius) shroud with smoothly distributed dust \citep{dudley+97,RR+09,roussel+06,siebenmorgen+08}.\footnote{
We note that the inner radius does not have to be at the dust sublimation radius, because the emergent SED is insensitive to the physical conditions within the $\sim5~\mu$m photosphere as long as the dust covering factor is unity.}
\cite{levenson+07} demonstrated, using IRAS~08572+3915 as an extreme example, that the very deep $9.7~\mu$m absorption requires such a geometrically thick dust shroud because it requires the large radial gradient in dust temperature.
NGC~4418 displays a similar but even more extreme SED (deeper $9.7~\mu$m absorption and redder MIR color\footnote{
The $9.7~\mu$m absorption depth: $S_{9.7}=-4.0$ for IRAS~08572+3915 \citep{levenson+07} and $-7$ for NGC~4418 \citep{roche+15}; the continuum flux ratio between $14~\mu$m (or $15~\mu$m) and $30~\mu$m: $S_{14}/S_{30}=0.2$ \citep{levenson+07} and $S_{15}/S_{30}=0.19$ \citep{spoon+22} for IRAS~08572+3915 and $S_{15}/S_{30}=0.11$ \citep{spoon+22} for NGC~4418.}), and such characteristics can be explained by more extreme shroud parameters according to the model of \cite{levenson+07}.\footnote{
An alternative idea involving an extended diffuse cold dust screen has been proposed to explain the deep $9.7~\mu$m absorption, e.g., by \cite{gm+13} and \cite{roche+15}.
However, we prefer the model involving a compact hot shroud, which is required to explain many other observed characteristics of this galaxy nucleus (e.g., \citealt{ga+12,costagliola+13,sakamoto+21,sakamoto+21b}), while the corresponding extremely large optical absorption ($A_\mathrm{V}>100$ mag) at the nucleus is not known \citep{scoville+00,evans+03,ohyama+19}.}
For NGC~4418, $r_\mathrm{sub}(\simeq r_\mathrm{in})\sim0.2$~pc assuming an AGN SED and dust sublimation temperature of 1500~K \citep{nenkova+08}, and the shroud extends out to $r_\mathrm{out}\gtrsim20$~pc well beyond the $5~\mu$m photosphere.
The gas and dust concentration around the nucleus actually extends out to $r=50$--$100$~pc, according to the ALMA molecular-gas observation of \cite{sakamoto+21b} and the FIR and submillimeter spectral modeling of \cite{ga+12}.
\cite{scoville13} and \cite{marshall+18} demonstrated that the dust shroud should not be clumpy; otherwise, characteristic dust spectral features are easily erased, because the emission from the inner hot dust or the hotter surface of the internal dust clumps contributes significantly to the emergent SED.
The large $C_\mathrm{f}$ suggests a large covering factor of the warm CO-absorbing region in front of the background light source (Section~\ref{sec:spectral_fitting}), and this can be naturally explained if the warm CO is located within, but in a very inner region of, such a dust shroud.
In contrast to the radially extended dust shroud, the CO-absorbing region is likely to be a thin layer, as
$dr_\mathrm{CO}=N_\mathrm{H2}/n_\mathrm{H2}/\phi_\mathrm{V}=5\times10^{23}$~cm$^{-2}/3\times10^6$~cm$^{-3}/\phi_\mathrm{V}\sim0.06/\phi_\mathrm{V}$~pc$=0.06$--$2$~pc,
where $\phi_\mathrm{V}$ (volume filling factor) is likely to be between 0.03 \citep{onishi+21} and 1.
We note that this radial size corresponds to the warm CO, which is only a fraction of the total CO and H$_2$.
The H$_2$ column density we measured ($N_\mathrm{H2}=(5\pm3)\times10^{23}$~cm$^{-2}$; Section~\ref{sec:rotation_diagram}) traces only this thin layer in front of the $5~\mu$m photosphere at $r\sim4$~pc (Section~\ref{sec:photosphere}) and, therefore, the total columns to the central heating source should be larger than our estimate, more significantly so for a thinner absorbing layer.

\subsection{Dynamical Implications for Warm CO Region} \label{sec:dynamical_implication}

The dynamical conditions of the CO-absorbing region measured with the $0\farcs04$ beam in front of the $5~\mu$m photosphere are characterized by both a broad width ($110$~km~s$^{-1}$ FWHM) and an absence of significant radial bulk motion along our line of sight ($|dV|<10$~km~s$^{-1}$; Section~\ref{sec:spectral_fit_result}).
A disk-like rotational velocity field around the nucleus contributes little to the observed CO absorption profile due to the relatively small projected velocity gradient ($\sim50$~km~s$^{-1}$ over $0\farcs2$; \citealt{sakamoto+21b}).
In addition, any radial velocity gradient within the absorbing layer is likely small if the layer is as thin as we estimated above (of the order of $0.1$--$1$~pc).
Therefore, we expect that the broad CO absorption profile is mainly due to a large turbulent motion in the CO-absorbing region.

\subsection{Comparison with Galaxies with Previous Rovibrational CO Measurements}\label{sec:comp_other_galaxies}

In an attempt to further characterize the circum-nuclear warm dense gas in NGC~4418, we compiled the rovibrational CO absorptions and other observed properties that are likely to similarly trace the warm dense gas, as we discuss in detail later, for a number of sources (Table~\ref{tab:tab2}).
To allow for a meaningful comparison across the sample, we utilize low-resolution ($R\sim100$) CO measurements with AKARI and Spitzer, along with the synthetic low-resolution spectrum based on the high-resolution measurement.
The AKARI spectroscopy is particularly efficient for detecting the CO absorptions in low-redshift galaxies due to its wavelength coverage below $5~\mu$m.
Our sample comes mainly from \cite{baba+22}, who compiled the AKARI CO measurements, and two others with other CO measurements.
We specifically compare galaxies with either deep CO absorption, enhanced vibrationally-excited HCN emission at submillimeter wavelengths, or both.
See also Table~\ref{tab:tab2} for the references.

\begin{deluxetable*}{lrllllcc}
\tablewidth{0pt} 
\tablecaption{Comparison of MIR Rovibrational Absorptions, HCN-vib Enhancement, and Other MIR Characteristics for Galaxies Displaying Rovibrational CO Absorption\label{tab:tab2}}
\tablehead{
\colhead{} & \colhead{$\tau(\mathrm{CO})$} & \colhead{C$_2$H$_2$ and HCN} & \colhead{HCN-vib} & \colhead{Spectral} & \colhead{Continuum} & \colhead{Refs} & \colhead{Refs} \\
\colhead{} & \colhead{} & \colhead{Absorption} & \colhead{Enhancement} & \colhead{Class} & \colhead{Slope} & \colhead{for (3)} & \colhead{for (6)} \\
\colhead{(1)} & \colhead{(2)} & \colhead{(3)} & \colhead{(4)} & \colhead{(5)} & \colhead{(6)} & \colhead{(7)} & \colhead{(8)}
}
\startdata
\multicolumn{8}{c}{AKARI Sample with CO Absorption of \cite{baba+22}}\\
\multicolumn{2}{c}{Deep CO Absorption}\\
IRAS 15250+3609 & $>1.0$ & Y & N & 3A & red ($>4~\mu$m) & 1 & 7\\
IRAS 08572+3915 NW & $0.7 $ & Y & N & 3A & red & 1 & 8\\
UGC 5101 & $0.7 $ & Y (shallow\tablenotemark{a}) & Y & 2B\tablenotemark{a} & red & 1 & 8\\
IRAS 20551-4250 & $>0.5$ & Y (C$_2$H$_2$\tablenotemark{b}) & N & 3B & red ($>3.5~\mu$m) & 2, 3, 4 & 7\\
NGC 4418 & $0.5 $ & Y & Y & 3A & red ($>4~\mu$m) & 1 & 7\\
\\
\multicolumn{2}{c}{Moderate CO Absorption}\\
Arp 220 & $0.4 $ & Y & Y & 3A & flat & 1 & 7\\
IRAS 17208-0014 & $0.4 $ & Y (shallow) & Y & 2B & flat & 1 & 9\\
\\
\multicolumn{2}{c}{Shallow CO Absorption}\\
IC 860 & $0.3 $ & Y & Y & 2B & blue & 1 & 7\\
Zw 049.057 & $0.3 $ & N & Y & 1C & blue & 5 & 7\\
\\
\hline
\multicolumn{8}{c}{Others}\\
IRAS F00183-7111 & $0.4$\tablenotemark{c} & \nodata\tablenotemark{d} & \nodata\tablenotemark{e} & 3B & red & 6 & 7, 10\\
NGC 4945 & $0.15$\tablenotemark{f} & \nodata & \nodata & 3A & red & \nodata & 11\\
\enddata

\tablecomments{Column 1: Object Names.
Column 2: AKARI CO optical depths from \cite{baba+22} in a decreasing order of the optical depths and others with known CO absorption properties (see the text).
Column 3: rovibrational C$_2$H$_2$ and HCN absorptions.
Column 4: enhancement of the HCN-vib luminosity relative to the total IR luminosity ($L_\mathrm{HCN-vib}/L_\mathrm{IR}>1\times10^{-8}$) from \cite{conquest1}.
Column 5: MIR spectral class of \cite{spoon+07} taken from \cite{spoon+22}.
Column 6: continuum slope at 3--$5~\mu$m in janskys.
Column 7: references to the rovibrational C$_2$H$_2$ and HCN absorptions (4).
Column 8: references to the continuum slope (6).
}

\tablenotetext{a}{Likely due to starburst contamination (see the text).}
\tablenotetext{b}{Deep C$_2$H$_2$ and marginal HCN absorptions are detected.}
\tablenotetext{c}{IRS measurement (see the text).}
\tablenotetext{d}{\cite{spoon+09} reported the upper limit.}
\tablenotetext{e}{\cite{imanishi+14} reported no detection without giving the upper limit.}
\tablenotetext{f}{Based on the model gaseous CO absorption spectrum of \cite{spoon+03} after removing the ice features (see the text).}

\tablerefs{1: \cite{lahuis+07}; 2: \cite{spoon+06}; 3: \cite{farrah+07}; 4: \cite{imanishi+16}; 5: \cite{ps+10}; 6: \cite{spoon+09}; 7: \cite{imanishi+10}; 8: \cite{imanishi+08}; 9: \cite{baba+22}; 10: \cite{baba+18}; 11: \cite{castro+14}.}

\end{deluxetable*}

\subsubsection{NGC~4945}\label{sec:N4945}

NGC~4945 hosts an obscured AGN although its MIR emission is dominated by starburst.
\cite{spoon+03} detected resolved gaseous rovibrational CO absorption lines (Section~\ref{sec:introduction}) as well as deep broad ice (mostly OCN$^-$ and polar CO ice) features around $\simeq4.7~\mu$m using a high-resolution ($R=3000$) ISAAC spectrum.
They derived $T_\mathrm{ex}=35^{+7.5}_{-2.5}$~K, log~$N_\mathrm{12CO}=18.3\pm0.1$~cm$^{-2}$, and FWHM$=50\pm5$~km s$^{-1}$, i.e., the CO-absorbing gas in NGC~4945 is much cooler, lower in column density, and less turbulent when compared to NGC~4418 and other galaxies displaying deep CO absorption within the sample of \cite{baba+22} (Section~\ref{sec:akari_co_sample}).
This result seems reasonable because the CO-absorbing gas in NGC~4945 is associated with an extended ($r=120$~pc) star-forming disk \citep{spoon+03}.

Although the low-resolution ($R\simeq160$) AKARI spectrum of NGC~4945 displays a deep absorption at $\simeq4.7~\mu$m \citep{castro+14}, it is mainly due to the OCN$^-$ and polar CO ice absorptions \cite{spoon+03} detected.
By using the best-fit parameters of the LTE model of \cite{spoon+03} and applying spectral smoothing to simulate the AKARI spectrum, we found $\tau(\mathrm{CO})$ to be $\simeq0.17$ after removing the contributions from these ice features.
Therefore, NGC~4945 displays only shallow gaseous CO absorption in the sample of \citealt{baba+22} (Section~\ref{sec:akari_co_sample}).

\subsubsection{IRAS~F00183-7111}

IRAS~F00183-7111 is a well-known ULIRG ($L_\mathrm{IR}\simeq7\times10^{12}~L_\mathrm{\sun}$; \citealt{spoon+04}) for its deep ice absorptions, but it also displays deep gaseous rovibrational CO absorption (Section~\ref{sec:introduction}).
\cite{spoon+04} presented its low-resolution ($R\simeq80$ at $5~\mu$m) Spitzer IRS spectrum and analyzed its blended CO absorption feature to find $T_\mathrm{ex}\sim720$~K and log~$N_\mathrm{CO}=19.5$~cm$^{-2}$.\footnote{
\cite{baba+18} also analyzed the same spectrum and derived $T_\mathrm{ex}\simeq330$~K and log~$N_\mathrm{CO}=21.2$~cm$^{-2}$.
Although their results appear different from those of \citealt{spoon+04}, the differences can be explained by degeneracy between $N_\mathrm{CO}$ and $T_\mathrm{ex}$ when analyzing the low-resolution spectra in general, as demonstrated by \cite{baba+18} (their Figure~12).}
They inferred an optically thick, warm, dense ($>3\times10^6$~cm$^{-3}$), and thin ($0.03$~pc) CO-absorbing layer in the vicinity of the nucleus.
Therefore, the CO-absorbing region in this ULIRG is similar to those in IRAS~08572+3915 (see below) and NGC~4418.

The apparent CO optical depth at the IRS resolution, measured in a very similar way to the AKARI CO measurement, is $\tau(\mathrm{CO, IRS})\simeq0.4$.
Because the blended absorption feature is relatively flat near its absorption peak mainly due to the high excitation temperature \citep{baba+18}, the CO optical depth at the AKARI resolution ($R\simeq160$) is expected to be similar to that at the IRS resolution.
Therefore, IRAS~F00183-7111 displays relatively deep CO absorption when compared to NGC~4418 and other galaxies displaying deep CO absorption in the sample of \cite{baba+22} (Section~\ref{sec:akari_co_sample}).
Unfortunately, neither vibrationally excited HCN emission at submillimeter wavelengths or absorptions of rovibrational C$_2$H$_2$ and HCN transitions around $14~\mu$m (see below) are detected \citep{spoon+09,imanishi+14}, making a meaningful comparison with the \cite{baba+22} sample difficult.

\subsubsection{IRAS~08572+3915}\label{sec:IRAS08}

IRAS~08572+3915 is the first ULIRG in which the rovibrational CO absorption lines were studied in detail using high-resolution spectroscopy (Section~\ref{sec:introduction}).
Although the physical properties of the CO-absorbing gas (e.g., excitation temperature, column density; \citealt{onishi+21}) in this galaxy are similar to those in NGC~4418, the kinematics and the excitation mechanism of the gas, as well as the inferred circum-nuclear structure, are quite different from those in NGC~4418.
\cite{onishi+21} found that each of the deep $^{12}$CO absorption lines shows a complex broad profile ($\sim300$~km~s$^{-1}$ in total; see also \citealt{shirahata+13}), and identified three kinematical components.
They argued that all of these components are associated with the clouds near the surface of the AGN torus along the line of sight of the observer at different radial distances from the AGN.
Each of these clouds displays either radial outflow or inflow motion along the line of sight.
The very warm ($\simeq720$~K) innermost component is in the LTE conditions, while the two outer components are likely radiatively excited.
The background light comes from hot ($\sim1500$~K) dust near the dust sublimation radius ($0.5$~pc).
They proposed an XDR heating model to explain the very large column of the warm CO-absorbing gas.
\cite{matsumoto+22} demonstrated that all of these characteristics can be reproduced by a hydrodynamic radiation-driven fountain model of AGN tori in low-luminosity AGNs.

Apart from the CO absorption, one notable difference between the two galaxies is that NGC~4418 displays enhanced rotational HCN emission at the vibrationally excited state ($v_2=1$) at submillimeter wavelengths (hereafter, HCN-vib emission), while IRAS~08572+3915 does not.
We will discuss its implications below using the AKARI sample with CO absorption.

\subsubsection{AKARI Sample with CO Absorption of \cite{baba+22}}\label{sec:akari_co_sample}

\cite{baba+22} compiled the published AKARI CO measurements for a sample of HCN-vib emission measurements, including NGC~4418 and IRAS~08572+3915, and investigated the conditions for deep rovibrational CO absorptions.
They suggested that deep CO absorptions are often found in objects with enhanced HCN-vib emission ($J=3$--$2$) relative to the infrared luminosity ($L_\mathrm{HCN-vib}/L_\mathrm{IR}>1\times10^{-8}$; \citealt{conquest1};
hereafter, we consider the HCN-vib emission to be enhanced when $L_\mathrm{HCN-vib}/L_\mathrm{IR}>1\times10^{-8}$).
For example, NGC~4418 is one of their sample galaxies with deep CO absorption (see below) and enhanced HCN-vib emission \citep{sakamoto+10,sakamoto+21b}, following the proposed trend of \cite{baba+22}.
\cite{baba+22} argued that the CO absorption is deep when the circum-nuclear conditions are similar to those expected in the CONs, which are defined to show $L_\mathrm{HCN-vib}/L_\mathrm{IR}>1\times10^{-8}$ \citep{conquest1}.\footnote{
\cite{conquest2} revised the CON definition to have larger surface HCN-vib brightness ($\Sigma_\mathrm{HCN-vib}$) than $1~L_\mathrm{\sun}~$pc$^{-2}$ and found that the CONs defined with large $\Sigma_\mathrm{HCN-vib}$ also display $L_\mathrm{HCN-vib}/L_\mathrm{IR}>1\times10^{-8}$.}
This is because HCN is vibrationally excited in the vicinity of compact MIR-emitting sources and the HCN-vib emission is enhanced when the HCN-vib-emitting gas encloses most of the nucleus before a wide-angle outflow develops and disrupts the obscuration (\citealt{conquest1,conquest2}; see also Section~\ref{sec:buried_N4418}).

Because the HCN-vib emission is the transition from the state that is almost exclusively radiatively excited by absorbing rovibrational HCN lines at $14.0~\mu$m (hereafter, rovibrational HCN absorption), the enhancement of the HCN-vib emission and the depth of the rovibrational HCN absorption are likely to be physically closely related (\citealt{sakamoto+10,ga+19,sakamoto+21b}).
In fact, NGC~4418 also displays the deep rovibrational HCN absorption as well as the rovibrational C$_2$H$_2$ ($13.7~\mu$m) absorption (\citealt{lahuis+07}; hereafter, rovibrational C$_2$H$_2$ absorption).
\cite{lahuis+07} argued that both absorptions originate from warm ($200$--$700$~K) dense ($\gtrsim1\times10^7$~cm$^{-3}$) gas, and that these absorptions likely trace the same gas as the CO absorptions.
For NGC~4418, \cite{lahuis+07} derived $T_\mathrm{ex,C2H2,HCN}=300$~K with an uncertainty of up to 30\% using a similar analysis as for the CO absorption on the rovibrational C$_2$H$_2$ and HCN absorptions, in rough agreement with our $T_\mathrm{ex,CO}$ measured with the CO rotation diagram (Section~\ref{sec:rotation_diagram}).
With such a possible physical relationship in mind, below we further characterize NGC~4418 by combining MIR information (rovibrational C$_2$H$_2$, HCN, and CO absorptions and spectral classification of \cite{spoon+07} that is based on the strengths of the PAH $6.2~\mu$m emission and the $9.7~\mu$m absorption) and the HCN-vib enhancement.

Five objects in the sample of \cite{baba+22} display deep CO absorptions ($\tau(\mathrm{CO})\gtrsim0.5$; Table~\ref{tab:tab2}).
All of them show characteristic red SEDs between $5$--$20~\mu$m without any sign of strong starburst contamination.
Among them, three objects (NGC~4418, IRAS~08572+3915 (Section~\ref{sec:IRAS08}), and IRAS~15250+3609) are classified as 3A (with the smallest PAH $6.2~\mu$m equivalent width and the deepest $9.7~\mu$m absorption), suggesting that warm dust emission from the compact region around the nucleus dominates their $5$--$20~\mu$m SEDs.
The other two objects are classified as 3B (with a slightly larger PAH $6.2~\mu$m equivalent width when compared to 3A; IRAS~20551-4250) and 2B (also with a slightly shallower $9.7~\mu$m absorption; UGC~5101).
These classes are often interpreted as a mixture of the starburst (1C) and the 3A SEDs (e.g., \citealt{spoon+07}).
In fact, UGC~5101 displays signs of starburst contamination at $8$--$15~\mu$m as revealed by the high-spatial-resolution ground-based observation of \cite{mp+15}.
Two (out of all five) objects (NGC~4418 and UGC~5101) display enhanced HCN-vib emission.
NGC~4418 also displays deep rovibrational C$_2$H$_2$ and HCN absorptions as mentioned above.
UGC~5101, on the other hand, displays moderate rovibrational C$_2$H$_2$ and HCN absorptions, and it is also likely due to the starburst contamination \citep{mp+15}.
It is noteworthy that the remaining three objects do not display enhanced HCN-vib emission but moderate to deep rovibrational C$_2$H$_2$ and HCN absorptions instead;
IRAS~15250+3609 and IRAS~08572+3915 display both deep rovibrational C$_2$H$_2$ and HCN absorptions, and IRAS~20551-4250 displays deep rovibrational C$_2$H$_2$ and marginal HCN absorptions.

Six objects in the sample of \cite{baba+22} display enhanced HCN-vib emission (Table~\ref{tab:tab2}).
Apart from NGC~4418 and UGC~5101 that display deep CO absorptions as mentioned above, two (out of four) objects display moderate $\tau(\mathrm{CO})$ ($=0.4$; Arp~220, IRAS~17208-0014).
Both are likely contaminated by starbursts because they are classified as 3B (for the whole eastern and western nuclei; Arp~220) and 2B (IRAS~17208-0014).
Another sign of such contamination is also found at $\lesssim5~\mu$m, where the continuum is almost flat and is much bluer than in the 3A objects, suggesting stellar continuum contribution.
This probably underestimates their CO depths.
Arp~220 displays deep rovibrational C$_2$H$_2$ and HCN absorptions, while IRAS~17208-0014 displays the shallower absorptions.
The remaining two objects (IC~860, Zw~049.057) display shallower CO absorptions ($\tau(\mathrm{CO})=0.3$).
They are classified as 1C (almost pure starburst; Zw~049.057) and 2B (between 3A and 1C; IC~860).
Their CO depths are likely severely underestimated due to the stellar continuum contamination, as suggested also by the much bluer continuum at $\lesssim5~\mu$m.
IC~860 displays deep rovibrational C$_2$H$_2$ and HCN absorptions despite the contamination.
Zw~049.057 displays little sign of rovibrational C$_2$H$_2$ and HCN absorptions most likely due to the severe starburst contamination (and perhaps also due to the lower S/N of the spectrum; \citealt{ps+10}).

In summary, we found that objects with deep rovibrational CO absorptions ($\tau(\mathrm{CO})\gtrsim0.5$; IRAS~15250+3609, IRAS~08572+3915 NW, UGC~5101, IRAS~20551-4250, and NGC~4418) often also display deep rovibrational C$_2$H$_2$ and HCN absorptions as well.
Objects with shallower CO absorptions ($\tau(\mathrm{CO})<0.5$) but with enhanced HCN-vib emission (Arp~220, IRAS~17208-0014, IC~860, and Zw~049.057) often display moderate to deep rovibrational C$_2$H$_2$ and HCN absorptions.
Their CO depths are likely underestimated due to contamination.
This explains, at least in part, the large scatter of $\tau(\mathrm{CO})$ for similarly enhanced HCN-vib emission.
Taking such a possible CO underestimation into account, the deep rovibrational CO absorption seems to be more closely associated with the deep rovibrational C$_2$H$_2$ and HCN absorptions than with the enhanced HCN-vib emission.

\subsection{The Buried Nucleus of NGC~4418}\label{sec:buried_N4418}

The close associations among the rovibrational C$_2$H$_2$, HCN, and CO absorptions above indicate that all of these absorption lines trace the warm gas along the line of sight toward the MIR-emitting compact source, whereas enhancement of the HCN-vib emission, which is powered by the MIR radiation (e.g., \citealt{sakamoto+10}), increases as the volume of the warm gas optically thick to the MIR emission increases \citep{ga+19}.
The HCN-vib is particularly enhanced when the circum-nuclear region is in a greenhouse condition (\citealt{ga+19}; see below).
The volume of such warm gas likely changes depending on whether the circum-nuclear obscuring structure is in the form of a torus with a large opening angle toward its pole directions or a fully enclosed shroud (\citealt{baba+22}; see their Figure~13 for a schematic illustration), or depending on the different evolutionary stages of the wide-angle outflow from the nucleus (before or after it disrupts the shroud; \citealt{conquest1}; see their Figure~2 for a schematic illustration; and references therein).
In this scenario, objects with both deep MIR rovibrational absorptions and large enhancement of the HCN-vib emission are likely to be buried in a compact dusty envelope of warm dense gas.
NGC~4418 most likely belongs to this class.
We note that NGC~4418 indeed exhibits large enough column density (log $N_\mathrm{H2}\gtrsim25$~cm$^{-2}$) to cause the HCN-vib enhancement \citep{sakamoto+21}.
IRAS~08572+3915, on the other hand, belongs to the case of an AGN torus seen from an almost edge-on viewing angle, as proposed by \cite{onishi+21} and \cite{matsumoto+22}.

As for NGC~4418, we argued that both warm dust and CO coexist within the compact nucleus despite considerable uncertainties in their radial distributions (Section~\ref{sec:where_CO}).
In such a situation, the so-called greenhouse effect will likely work efficiently \citep{ga+19,sakamoto+21}.
In contrast to typical PDR/XDR models (e.g., \citealt{meijerink+05}), gas cooling in greenhouse is inefficient due to high opacity at MIR and FIR, and CO can be efficiently heated over a larger volume by gas--dust collisions due to higher gas density and hotter dust temperature.
This mechanism creates an extended warm CO region beyond the limits of the typical models (up to $N_\mathrm{CO}\sim10^{16}$~cm$^{-2}$ and $\sim10^{18}$~cm$^{-2}$ for warm, $>100$~K, CO in PDR and XDR, respectively; \citealt{baba+18}).
To test the greenhouse effect for NGC~4418, we need a better understanding of the dust distribution within the presumed $5~\mu$m photosphere, in particular, whether the dust shroud fully encloses the central heating source (or how efficient the greenhouse effect is) and whether its covering factor is unity (or whether the shroud has small holes through which observers can see inside).

To obtain more insights into the compact dusty gaseous shroud around the central heating source, we need a self-consistent modeling of the entire gas, dust, and radiation field in a more realistic environment for NGC~4418 to simultaneously reproduce the CO excitation and its absorption, the deep $9.7~\mu$m absorption, and the steep drop of the SED below $\sim5~\mu$m.
Our analysis of the warm CO absorptions (Sections~\ref{sec:spectral_fitting}, \ref{sec:spectral_fit_result}) assumed a simple foreground layer, despite that the warm CO is most likely embedded in a geometrically thick dust shroud with the radial gradients in temperature and density.
Ideally, such a model should also include FIR and submillimeter information \citep{ga+12,costagliola+13,mashian+15,sakamoto+21,sakamoto+21b}.
In particular, the deep (with an apparent optical depth of $\sim0.9$) and broad (spanning over 250~km~s$^{-1}$ in total including its multiple unresolved transitions spanning over 126~km~s$^{-1}$) CN ($J=6$--$5$) absorption at 680~GHz measured with a $0\farcs15$ beam toward the center \citep{sakamoto+21b} is noteworthy for its similarity to the rovibrational CO absorptions in terms of both the depth and the line width.
The CN lines are often used as tracers of highly excited regions and, in principle, can explore even deeper than the CO absorptions due to their lower opacity at submillimeter wavelengths than at MIR.

\section{Summary}

We investigated the buried nucleus of the nearby LIRG NGC~4418 using fundamental CO rovibrational absorptions and inferred a large column density ($N_\mathrm{H2}\sim5\times10^{23}$~cm$^{-2}$ in front of the $5~\mu$m photosphere) of warm ($T_\mathrm{kin}\simeq T_\mathrm{ex}\simeq170$~K) molecular gas by assuming an isothermal plane-parallel slab illuminated by a compact background MIR-emitting source.
The very deep and partly saturated $^{12}$CO absorptions indicate the large covering factor ($>0.86$).
The absorption profiles are broad ($110$~km~s$^{-1}$ FWHM) near the systemic velocity ($|dV|<10$~km~s$^{-1}$), suggesting a large turbulent motion with little bulk radial motion within the warm CO gas.
We modeled that the warm CO absorber almost covers the central heating source and that it is an inner thin layer (of the order of $0.1$--$1$~pc) around the $5~\mu$m photosphere (at $r=$several parsecs) of a compact shroud of gas and dust ($d\sim100$~pc).

\begin{acknowledgments}
We thank Drs. T. Usuda and S. Oyabu for their support in preparing and performing the observations.
Y.O. and K.S. acknowledge the support from the Ministry of Science and Technology (MOST) of Taiwan through the grants MOST 109-2112-M-001-021- (Y.O.) and MOST 111-2112-M-001-039- (K.S.).
T.N. and S.B. are supported by JSPS KAKENHI grant Nos. 21H04496 (T.N. and S.B.) and 23H05441 (T.N.).
K.M. is a Ph.D. Fellow of the Flemish Fund for Scientific Research (FWO-Vlaanderen) and acknowledges the financial support provided through grant No. 1169822N.
\end{acknowledgments}

\vspace{5mm}
\facilities{Subaru(IRCS)}

\software{emcee \citep{emcee}}

\bibliography{N4418_MIR_COABS_arxiv}{}
\bibliographystyle{aasjournal}

\appendix

\section{Spectral Fitting with the LTE Model}\label{appendix:LTE_fit}

In Section~\ref{sec:spectral_fitting}, we modeled the spectrum with all of the optical depths of individual $^{12}$CO and $^{13}$CO absorption lines as free parameters to fit (the full model) and then fitted the column densities of various $J$th rotational states on the rotation diagrams to obtain the excitation temperatures and the column densities (Section~\ref{sec:rotation_diagram}).
With this approach, we obtained a good reduced-$\chi^2$ ($\simeq1.1$ with 1985 degrees of freedom) in the spectral fitting (Figure~\ref{fig:fig1}; Section~\ref{sec:spectral_fit_result}) and found the level populations that are similar to what is expected in the LTE conditions (Figure~\ref{fig:fig2}; Section~\ref{sec:rotation_diagram}).
A main concern in this approach is the robustness of the results;
this model includes as many as 55 free parameters to fit the complicated spectrum without imposing any physical constraints on 47 absorption lines of $^{12}$CO and $^{13}$CO across wide ranges of $J$s, from which 26 lines were selected for the analysis with the rotation diagrams.

\begin{figure*}
\centering
\epsscale{1.15}
\plotone{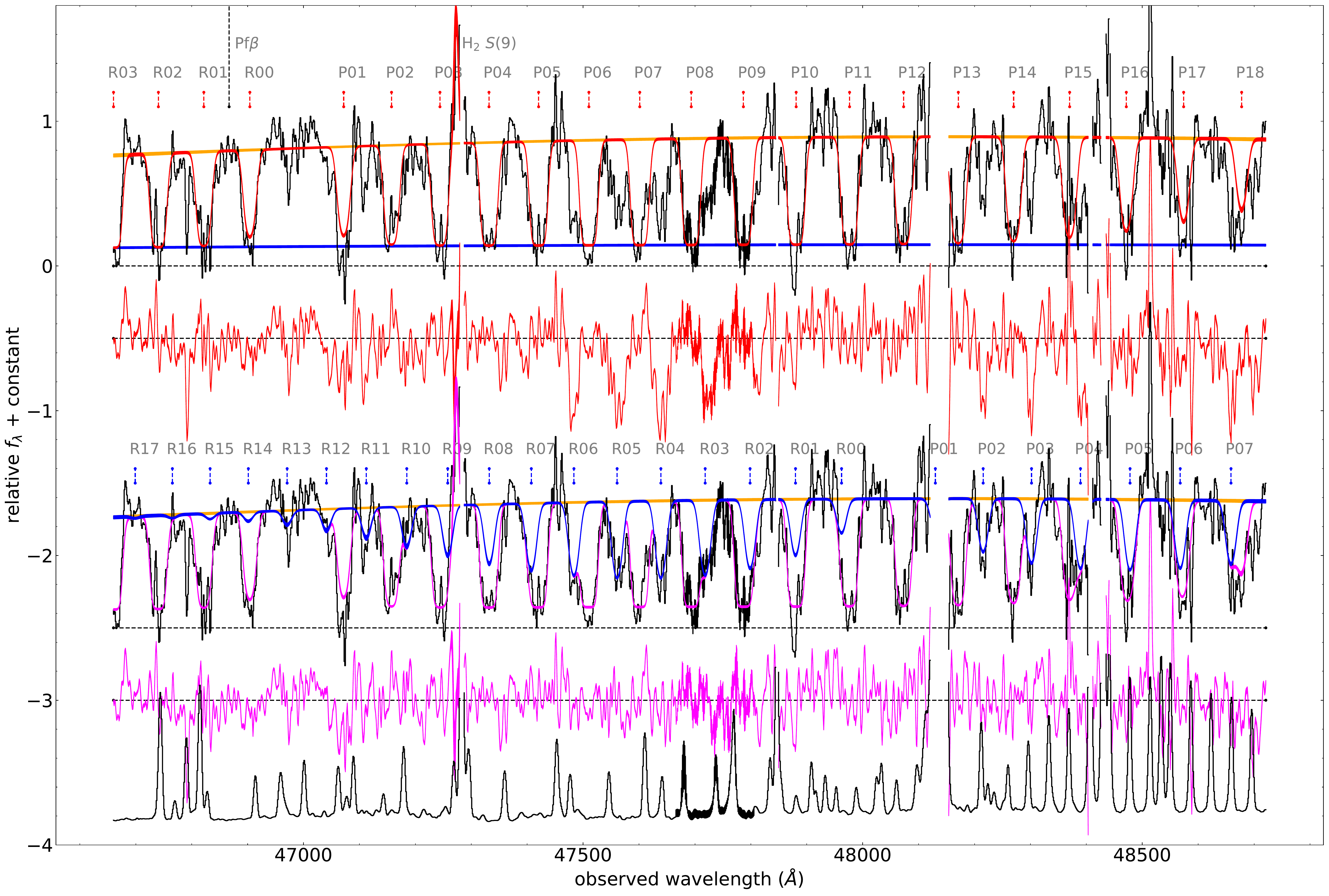}
\caption{
Similar spectral fit results as in Figure~\ref{fig:fig1}, but with the LTE model.
}
\label{fig:figa1}
\end{figure*}

To test the robustness of the fit, we also modeled the spectrum by assuming LTE conditions (the LTE model; Section~\ref{sec:spectral_fitting}).
The total number of free parameters is now reduced to 12.
We found a similarly good, but slightly worse, reduced-$\chi^2$ ($\simeq1.2$ with 2028 degrees of freedom; Figure~\ref{fig:figa1}) when compared to the full model results.
The level populations on the rotation diagram predicted with the LTE model are very similar to those we measured with the full model (Figure~\ref{fig:fig2}).
We derived $T_\mathrm{LTE,12CO}=260\pm8$~K and $T_\mathrm{LTE,13CO}=137\pm8$~K, and the total column densities $N_\mathrm{LTE,12CO}=(1.9\pm0.1)\times10^{19}$~cm$^{-2}$ and $N_\mathrm{LTE,13CO}=(3.6\pm0.2)\times10^{18}$~cm$^{-2}$, and found that all of these parameters are very similar to those with the full model (Section~\ref{sec:rotation_diagram}).
This confirms the robustness of the full model results, which we adopted in the main text.

On a closer look, we found small systematic differences between the best-fit full and the LTE models.
The LTE model slightly underestimates the depth of the $P(1)$ $^{12}$CO absorption, whereas the full model does not, on the spectrum (Figures~\ref{fig:fig1} and \ref{fig:figa1}).
The corresponding deviation is also found on the rotation diagram (Figure~\ref{fig:fig2});
the LTE model expects a smaller $^{12}$CO $P(1)$ column when compared to the full model measurement.
Such a difference can be caused by an additional lower-temperature component that dominates the low-$J$ populations, as \cite{shirahata+13} proposed for IRAS~08572+3915.
It is beyond the scope of this paper to improve the LTE model fit by adding more components.
The LTE model also expects a smaller $R(16)$ $^{13}$CO column when compared to the full model measurement.
This difference seems coupled to the systematically lower continuum with the LTE model when compared to that with the full model near the blue end of the wavelength coverage.
Without proper placement of the continuum, as done by, e.g., \cite{baba+18}, using low-resolution spectra with wider spectral coverage, it seems difficult to find a better model.

\end{document}